\documentclass[12pt]{article}
\usepackage{fullpage,amsfonts,latexsym,amsmath,epsf,latexsym,amssymb}
 
\newtheorem{definition}{Definition}
\newtheorem{theorem}{Theorem}
\newtheorem{lemma}{Lemma}

\newtheorem{remark}{Remark}

\newcommand{\R}[0]{{\mathbb{R}}}

\newcommand{\Z}[0]{{\mathbb{Z}}}

\def\C{{\mathbb{C}}}
\def\Z{{\mathbb{Z}}}
\def\R{{\mathbb{R}}}

\def\N{{\mathbb{N}}}

\newcommand{\cH}{{\cal H}}

\newcommand{\cO}{{\cal O}}

\newcommand{\proof}[1]{{\bf Proof.} #1 \hfill $\Box$\vspace{0.5cm}}

\newcommand{\ket}[1]{|#1\rangle}
\newcommand{\bra}[1]{\langle #1|}

\title{Estimating mixing properties of local Hamiltonian dynamics and
continuous quantum random walks is PSPACE-hard}

\author{Pawel Wocjan\thanks{e-mail: {\protect\tt
wocjan@ira.uka.de}\quad I'm looking for a postdoc position. Please
email me if you could suggest me something.}\\ \small Institut f{\"u}r
Algorithmen und Kognitive Systeme, Universit{\"a}t Karlsruhe,\\[-1ex]
\small Am Fasanengarten 5, D-76\,131 Karlsruhe, Germany}

\date{January 27, 2004}

\begin{document}

\maketitle

\abstract{A major topic of (classical) ergodic theory is to examine
qualitatively how the phase space of dynamical systems is penetrated
by the orbits of their dynamics. We consider interacting qubit systems
with dynamics according to $4$-local Hamiltonians and continuous
quantum random walks. For these systems one could use the von Neumann
entropy of the time-average to characterize the mixing properties of
the corresponding orbits, i.e., what portion of the state space and
how uniformly it is filled out by the orbits. We show that the problem
of estimating this entropy is PSPACE-hard.}

\section{Introduction}
Ergodic theory studies the long-term average behavior of dynamical
systems. The continuous time-evolution is described by a one-parameter
family $\{T_t\, :\, t\in\R\}$ of maps of the phase space $X$ into
itself. A major topic of classical ergodic theory is to examine
qualitatively how the orbits $\{T_t x\}$ (where $x$ is the initial
state) penetrate the phase space $X$. If the orbits ``spread
uniformly'' over $X$, then the system is called ergodic. Ergodicity of
dynamical systems plays an important role in statistical mechanics and
thermodynamics because in the ergodic case the time averages and
ensemble averages of physical variables are equal (see
\cite{Petersen:83} for a rigorous study of these questions).

In this paper we consider the case that $X$ is the Hilbert space of
many qubits, and $\{T_t : t \in\R\}$ is the unitary one-parameter
group $\{e^{-i H t} : t\in\R\}$ generated by the system Hamiltonian
$H$. We study the computational complexity of estimating the von
Neumann entropy generated by Hamiltonian time evolutions if we average
over the time. The motivation for considering the entropy of the time
average is that it quantifies in a certain sense what portion of the
state space and how uniformly it is filled out by the trajectory
according to the time evolution. A second interesting property of the
entropy of the time-average is that it could be used to quantify how
far a pure state $\ket{\Phi}$ is from all eigenvectors of a
Hamiltonian (this question was raised in \cite{HNO:03}). The reason is
that the entropy of the time-average of the initial pure state
$\ket{\Phi}\bra{\Phi}$ can only be large if the initial state is a
superposition of many eigenvectors (corresponding to different
eigenvalues) of $H$ with equal amplitudes. We explain this later in
more detail.

The time-average is defined as follows:
\begin{definition}[Time average]${}$\\
Let $H$ be the Hamiltonian of a quantum system $\cH$. For an arbitrary
state $\rho$ on $\cH$ its time average is the state $\bar{\rho}$
defined by
\begin{equation}
\bar{\rho} := \lim_{T\rightarrow\infty} 
\frac{1}{T}
\int_{t=0}^T e^{-i H t}\,\rho\, e^{i H t}\, {\rm d} t\,.
\end{equation}
\end{definition}
It is well-known that the time average depends on the spectral
decomposition of the Hamiltonian (a proof of the following lemma is
included in the appendix for completeness).
\begin{lemma} The time-average $\bar{\rho}$ of $\rho$ is given by
\[
\bar{\rho}=\sum_{\lambda} P_{\lambda} \rho P_{\lambda}\,,
\]
where $P_{\lambda}$ are the orthogonal projections onto the
eigenspaces of $H$ for different eigenvalues $\lambda$.
\end{lemma}
Before we can formulate question on the complexity of estimating
certain properties of time averages we have to restrict the class of
considered Hamiltonians. This is because physically realistic
Hamiltonians must satisfy some locality condition. One way to
formalize this is the following definition.
\begin{definition}[local operator]${}$\\
Let $\cH:=\cH_1\otimes\cdots\otimes \cH_n$ be the tensor product
Hilbert space of $n$ Hilbert spaces. We call an operator $k$-local if
it is a sum of operators that act on at most $k$ tensor components
non-trivially.
\end{definition}
If each tensor component is the state space of a physical particle,
then the natural interactions are usually pair-interactions, that is,
$2$-local operators in the sense of the above definition.

Nevertheless, for the Hilbert space $\cH:=(\C^2)^{\otimes n}$ of $n$
interacting qubits it may also be physically reasonable to allow
$k$-local interaction for $k>2$. One reason is that they may describe
effective Hamiltonians. A second reason is that there is not
necessarily a one-to-one correspondence between qubits and physical
particles (for instance, several qubits could describe the state of
one particle).

We consider time-evolutions according to $4$-local Hamiltonians and
continuous quantum random walks. The time evolution $e^{-i H t}$ is
called a continuous quantum random walk if $H$ is the adjacency matrix
of some graph \cite{CCD:02}. Recall that a matrix $A$ is called an
adjacency matrix if it is symmetric, has only $0$s and $1$s as
entries, and all its entries on the diagonal are $0$. Furthermore, we
say that a matrix is a $k$-local adjacency matrix if it is a sum of
$k$-local operators that are adjacency matrices themselves. We will
consider $5$-local adjacency matrices.

We show that the problem estimating the von-Neumann entropy of
time-averages of computational basis states, where the time-evolution
is described by $4$-local Hamiltonians and $5$-local quantum random
walks, is PSPACE hard. The problem of estimating the density matrix of
the time-average reduced to a qubit is also PSPACE-hard. 

Due to Lemma~1 it is clear that the entropy generation depends on the
whole spectrum of the Hamiltonians. Therefore, these results should be
compared to the results on the complexity of estimating the minimal
eigenvalue (ground state energy) of local Hamiltonians. The problem of
determining the lowest energy value of a (classical) spin-spin
interaction of Ising type is known to be NP-complete
\cite{Barahona:82,Pawelcompass}. For interacting qubits the problem of
determining the lowest energy value is even QMA-complete
(``Quantum-NP'') if one allows $3$-local interactions only
\cite{KitaevShen,KempeRegev}. Recall that in these NP and Quantum-NP
problems the task is not to determine the lowest eigenvalues {\it with
high precision}. The demanded accuracy is only inverse polynomially in
the number $n$ of interacting qubits.

Note that the entropy of the time-average of a pure state
$\rho:=\ket{\Phi}\bra{\Phi}$ could be used to measure how ``far away''
$\ket{\Phi}$ is from all eigenstates of $H$. This question was studied
in \cite{HNO:03}; the authors constructed quantum states that are far
away from any eigenstate of any non-trivial local Hamiltonian, in the
sense that $\|\ket{\Phi}-\ket{\Psi}\|$ is greater than some constant
lower bound for all eigenstates $\ket{\Psi}$ of all $k$-local (for a
fixed $k$) Hamiltonians, independent of the form of the Hamiltonians.
The measure for the distance of a pure quantum state $\ket{\Phi}$ from
the spectrum of a Hamiltonian used in \cite{HNO:03} is 
\[
d(\ket{\Phi},H):=
\min_{\{\ket{\Psi_{\lambda}}\}} \min_{\lambda}
\|\ket{\Phi} - \ket{\Psi_\lambda}\|\,,
\]
where $\{\ket{\Psi_\lambda}\}$ runs over all sets consisting of
eigenvectors of $H$ corresponding to different eigenvalues
$\lambda$. One easily verifies that the distance is equivalent to
\[
\min_{\{\ket{\Psi_{\lambda}}\}} \min_{\lambda}
2(1-|\alpha_{\lambda}|^2)\,,
\]
where $\ket{\Phi}=\sum_{\lambda} \alpha_{\lambda}
\ket{\Psi_{\lambda}}$. Therefore, the bound $d(\ket{\Phi},H)$ is
maximal if and only if $\ket{\Phi}$ is an equally weighted
superposition of eigenvectors corresponding to different eigenvalues of
$H$. Using Lemma~1 one see that this is equivalent that the entropy of
the time average of $\ket{\Phi}$ is maximal; the maximally possible
entropy is the logarithm of the number of different eigenvalues of
$H$. Based on these observations, our results may be also
interpreted as a proof of the computational difficulty of deciding for
a given quantum state how far it is from all eigenstates of a local
Hamiltonian.

\section{Stating the problem}
The techniques we use here to prove PSPACE-hardness of estimating the
entropy of time-averages are closed related to those in \cite{WJDB:03}
where we studied the complexity of measuring local observables. The
general idea is to describe the computational steps carried out by a
(reversible) space-bounded Turing machine by a quantum circuit
consisting of elementary quantum gates or Toffoli gates and then to
encode the circuit into a local Hamiltonian (this will be explained in
Sections $3$ and $4$).

\begin{theorem}[PSPACE-hardness]${}$\\
Let $(A_l)$ be a uniformly generated family of $4$-local Hamiltonians
(or $5$-local adjacency matrices) on the Hilbert spaces $H_l$
consisting of $poly(l)$ qubits. Then the problems of (1) estimating
the von-Neumann entropy of the time-average $S(\bar{\rho})$ and (2)
estimating the reduced state of $\bar{\rho}$ to single qubits are
PSPACE-hard.

More precisely: (1) Deciding if the entropy of the time-average of a
computational basis state of $\cH_l$ according the dynamics given by
$A_l$ is smaller than $a_l$ or greater than $b_l$, where
$b_l-a_l>poly(l)$, is PSPACE-hard.

Let $\bar{\rho}_i$ denote the reduced state of the time-average
$\bar{\rho}$ to the $i$th qubit of $\cH_l$. (2) Deciding if
$\bra{1}\bar{\rho}_i\ket{1}<1/d_l$ or
$\bra{1}\bar{\rho}_i\ket{1}>1/2-1/d_l$, where $d_l=poly(l)$, is
PSPACE-hard.
\end{theorem}
We prove this theorem in Section~4. The proof is based on a
characterization of the complexity class PSPACE with quantum circuits
that is explained in the next section.

\section{Characterizing PSPACE by circuits} 
The complexity class PSPACE is usually defined with respect to the
Turing machine model \cite{HU:79}. PSPACE is the class of all
languages recognizable by polynomial space bounded deterministic
Turing machines that halt on all inputs \cite{GJ}.

In the following we work with a characterization of PSPACE with
respect to circuits consisting of elementary quantum gates (Toffoli
gates only). This characterization is proved in Theorem~1 in
\cite{WJDB:03}.

\begin{figure}
\centerline{\epsfxsize0.75\textwidth\epsfbox{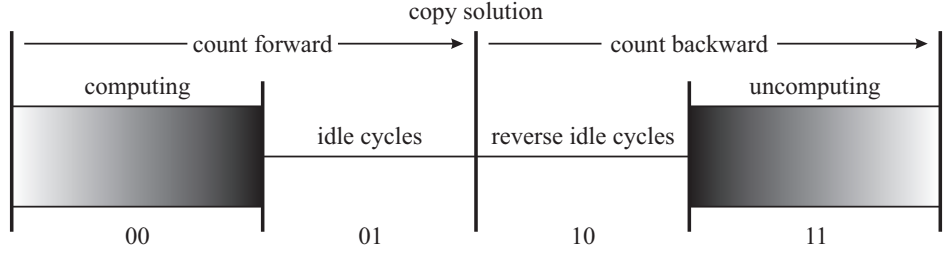}}
\label{schema}
\caption{Schematic representation of how $V$ works}
\label{fig:schema}
\end{figure}

\begin{theorem}[PSPACE]\label{th:pspace}${}$\\
For every language $L$ in PSPACE there is a polynomial-time uniformly
generated family of quantum circuits $(V_l)_{l\in \N}$. Here $l$
denotes the length of the input $x$. Each $V_l$ consists of
$s_l=poly(l)$ elementary quantum gates (or Toffoli gates\footnote{In
\cite{WJDB:03} we have used elementary quantum gates instead of
Toffoli gates. But it is clear that the transitions performed
reversible Turing machines can be simulated by a circuit consisting of
Toffoli gates only. This is because the Toffoli gate is universal for
reversible computation. The reason why we use the Toffoli gates is
that it is possible to construct from such a circuit a Hamiltonian is
an adjacency matrix.}) and acts on $m_l=poly(l)$ many qubits. The
circuit $V_l$ decides whether an input string $x$ is an element of $L$
in the following sense.

There is a polynomial-time computable natural number $r_l$ such that
the $r_l$-fold concatenation of $V_l$ solves the corresponding PSPACE
problem, i.e.\
\[
V_l^{r_l} (\ket{x} \otimes \ket{y} \otimes \ket{00\dots 0}) =
|x\rangle \otimes |y \oplus f(x)\rangle  \otimes |00\dots 0\rangle\,,
\]
where $f$ is the characteristic function of $L$. That is $f(x)=1$ if
$x\in L$ and $f(x)=0$ otherwise. The vector $|x\rangle$ is the basis
state given by the binary word $x \in \{0,1\}^l$, the vector
$|y\rangle$ is the state of the output qubit and $|00\dots 0\rangle$
is the initial state of $m_l-l-1$ ancilla qubits.
\end{theorem}
The intuition behind this construction is as follows. Recall that
irreversible TMs (used to define PSPACE) can be simulated
space-efficiently by reversible TMs \cite{Bennett:89,LMT:00}. The
computational steps of a reversible TM can be simulated efficiently by
a quantum circuit $U$. This circuit acts on registers representing the
state of the head of TM, the current tape position, and a polynomial
portion of the tape (see \cite{WJDB:03} for an explicit construction
of the quantum circuit $U$) and realizes permutations of basis states
corresponding to the transitions of the TM. By augmenting $U$ with
some control logic, we can achieve that once the TM moves into a final
state, that the new circuit performs some idle cycles, flips the
output qubit if and only if the answer is yes after exactly $r_l/2$
steps, then performs some idle cycles, and reverses the
computation. This is schematically illustrated in
Figure~\ref{fig:schema}. It is necessary to perform idle cycles
because we sometime do not know exactly after how many steps the TM
moves into a final step, but only know a lower bound. By including the
idle steps we guarantee the computational time depends only on the
input length.

\begin{remark}
We can consider an output register instead of an output qubit.  It is
also possible to construct a quantum circuit such that
\[
V_l^{r_l} (\ket{x} \otimes \ket{y} \otimes \ket{00\dots 0}) =
|x\rangle \otimes |y + f(x)\!\! \mod 2^c\rangle  \otimes |00\dots 0\rangle\,,
\]
for any $c$ that is polynomial in $l$.
\end{remark}

\section{Constructing the Hamiltonians\,/\\ the quantum random walks}
Starting from the family $(V_l)$ of circuits we construct a uniformly
generated family of $4$-local Hamiltonians ($5$-local adjacency
matrices) $(A_l)_{l\in\N}$ such that the time-average of the state
\[
\ket{x}\otimes\ket{0}\otimes\ket{00\ldots 0}\otimes\ket{100\ldots 0}
\]
and its entropy encodes the solution the instance $x$ of a PSPACE
problem; the first three tensor components are as in Theorem~2 and the
last one corresponds to the new register \texttt{clock} that is needed
in our construction. The construction is based on Feynman's
construction of a computer whose dynamics is an autonomous
time-evolution \cite{Feynman:85,Margolus:90}.

To explain the construction of the Hamiltonian (adjacency matrix $A$)
let $V$ be a circuit as in Theorem~2 and $s$ be its size $s$, that is,
the number of elementary gates (Toffoli gates). We need a register
\texttt{clock} indicating which gate is applied. It consists of $s$
qubits. The allowed states of the register \texttt{clock} are of the
form $ \ket{0\cdots 0 1 0 \cdots 0} $ indicating which gate of $V$ is
applied currently. We denote by $T_k$ the elementary quantum gates
(Toffoli gates) of $V$.

We first define the {\em forward-time} operator
\begin{eqnarray*}
F & = &
T_0 \otimes \ket{1}_{2}\bra{0}_{1} \otimes \ket{0}_{1}\bra{1}_{0} + \\
& &
T_1 \otimes \ket{1}_{3}\bra{0}_{2} \otimes \ket{0}_{2}\bra{1}_{1} + \\
& &
\,\vdots \\
& &
T_{s-1} \otimes \ket{1}_{1}\bra{0}_{0} \otimes \ket{0}_{s}\bra{1}_{s-1}
\,,
\end{eqnarray*}
where the operators $|0\rangle_k \langle 1|_k$ and $|1\rangle_k
\langle 0|_k$ are the annihilation $a$ and creation operators
$a^\dagger$, respectively, acting on the $i$th qubit of the
\texttt{clock}. The {\em backward-time} operator is defined as the
adjoint of $F$. The {\em Hamiltonian} is defined as
\[
A:=F+F^\dagger\,.
\]
Now we show that if $V$ consists of Toffoli gates only, then $A$ is a
$5$-local adjacency matrix of some graph in the computational basis,
that is, $A$ is symmetric, has only $0$s and $1$s as entries, and its
entries on the diagonal are all $0$s. Since the matrix describing the
action of the Toffoli gate (in the computational basis) contains only
$0$s and $1$s, the sums
\[
T_k \otimes \ket{1}_{k}\bra{0}_{k} \otimes 
            \ket{0}_{k+1}\bra{1}_{k+1} +
T_k \otimes \ket{0}_{k}\bra{1}_{k} \otimes 
            \ket{1}_{k+1}\bra{0}_{k+1}
\]
are adjacency matrices that are $5$-local operators. $A$ is the sum of
the above matrices. Therefore, it is symmetric and has only $0$s on
the diagonal. It remains to show that no $1$s of these matrices can
meet when summing. This is done by checking that the operators
\[
a_k^\dagger a_{k+1 \!\!\mod s}\,,\,
a_k a_{k+1 \!\!\mod s}^\dagger\,,
\]
lead to orthogonal states for all $k=0,\ldots,s-1$ when applied to any
computational basis state. We see this by observing that if $k-l \mod
s\ge 2$ then we have
\[
\bra{b} a_k^\dagger a_{k+1}\, a_l a_{l+1}^\dagger\ket{b}=0
\quad\mbox{ and }\quad
\bra{b} a_k a_{k+1}^\dagger\, a_l a_{l+1}^\dagger\ket{b}=0
\] 
because the operators $a_k^\dagger a_{k+1}\, a_l a_{l+1}^\dagger$ and
$a_k a_{k+1}^\dagger\, a_l a_{l+1}^\dagger$ either map $\ket{b}$ onto
the zero vector or change the bits of $\ket{b}$ at the non-overlapping
positions $(k,k+1)$ and $(l,l+1)$. Similarly, if $l=k+1$ then we have
\[
\bra{b} a_k^\dagger a_{k+1}\,         a_{k+1} a_{k+2}^\dagger\ket{b}=0
\quad\mbox{ and }\quad
\bra{b} a_k         a_{k+1}^\dagger\, a_{k+1} a_{k+2}^\dagger\ket{b}=0
\] 
because $a_k^\dagger a_{k+1}\,a_{k+1} a_{k+2}^\dagger$ is the zero
operator and $a_k a_{k+1}^\dagger\, a_{k+1} a_{k+2}^\dagger$ either
maps $\ket{b}$ onto the zero vector or changes the bits of $\ket{b}$
at the non-overlapping positions $k$ and $k+2$.

Therefore, only computational basis states appear in $A\ket{b}$ with
coefficients all equal to $1$. Consequently, we have either $\bra{b} A
\ket{b}=0$ or $\bra{b} A \ket{b}=1$. This show that $A$ is a $5$-local
adjacency matrix if $V$ contains Toffoli gates only.

We denote the linear span of the vectors
\[
F^j|\Psi_0\rangle \quad\mbox{for } j\in \N
\]
where $\ket{\Psi_0} := \ket{x\, 00\cdots 0} \otimes\ket{100\cdots 0}$
as $\cO$. Here the first part of the tensor product denotes the
register where $V$ acts on (input and ancilla registers) and the
second component is the register \texttt{clock}.  All states of this
orbit are orthogonal until one has a recurrence to the initial state
$|\Psi_0\rangle$. This can be seen as follows: If the register
\texttt{clock} is in an allowed state there is only one summand of $F$
that is relevant. Its action on the \texttt{clock} is simple since it
moves the symbol $1$ to the next qubit (this dynamics of the
\texttt{clock} may be interpreted as a propagation of a spin-wave).
Therefore it is clear that the first $s-1$ states are orthogonal. The
whole circuit $V$ is a classical logical operation which permutes
basis states. Therefore the state $F^s |\Psi_0\rangle$ is either
orthogonal to $|\Psi_0\rangle$ or both states coincide. Along the same
line we argue that all states of the orbit are orthogonal until a
state coincides with the initial state. Hence $F$ acts as a cyclic
shift on $\cO$.

For our construction is is essential that the dimension of the orbit
depends on the solution of the PSPACE problem. It is $2 s r$ if
$f(x)=1$ and $s r$ if $f(x)=0$, where $f(x)$ is the answer for the
instance $x$ of the considered PSPACE problem. We denote the dimension
of the orbit $\cO$ by $d$. Let $\omega$ be a primitive complex $d$-th
root of unity. The eigenvalues of $F$ restricted to $\cO$ are
\[
\omega^0,\omega^1,\omega_2,\ldots,\omega^{d-1}\,.
\]
Furthermore, $\ket{\Psi_0}$ is a superposition of all eigenvectors of
$F$ restricted to $\cO$ with equal amplitudes. All this follows from
properties of the cyclic shift operator. Since $F$ and $F^\dagger$
commute on $\cO$ the eigenvalues of $A$ restricted to $\cO$ are
$\omega^k+\bar{\omega}^k=2\cos(2\pi j/d)$ for $k=0,\ldots,d-1$.

Set $\rho:=\ket{\Psi_0}\bra{\Psi_0}$ and $\lambda_k:=2\cos(2\pi
k/d)$. Note that $d$ is always even; this is because $r$ in Theorem~2
is even.  Therefore, there are two eigenvalues with multiplicity $1$
(corresponding to the real eigenvalues $+1$ and $-1$ of $F$) and $d-2$
eigenvalues with multiplicity $2$ (corresponding to the complex
eigenvalues of $F$). We have
\[
\rho=\frac{1}{d}\sum_{k,l=0}^{d-1} \ket{k}\bra{l}\,,
\]
where $\ket{k}$ denotes the $k$th eigenvalue of $F$ restricted to
$\cO$. For the time-average we obtain with Lemma~1
\[
\bar{\rho} =
\frac{1}{d} \sum_{k,l=0}^{d-1} [\lambda_k=\lambda_l]\, \ket{k}\bra{l}\,,
\]
where $[\lambda_k=\lambda_l]$ is $1$ if both eigenvalues are equal and
$0$ otherwise. If we permute the order of the eigenvectors $\ket{k}$
such that eigenvectors corresponding to the same eigenvalues are
adjacent, then $\rho$ looks in this basis as follows:
\begin{equation}\label{eq:timeaverage}
\frac{1}{d}
\left(
\begin{array}{ccccccc}
1 &   &   &   &        &   &   \\
  & 1 &   &   &        &   &   \\
  &   & 1 & 1 &        &   &   \\
  &   & 1 & 1 &        &   &   \\
  &   &   &   & \ddots &   &   \\
  &   &   &   &        & 1 & 1 \\
  &   &   &   &        & 1 & 1 \\
\end{array}
\right)
\end{equation}
Using the form of the time-average in (\ref{eq:timeaverage}), its
entropy is easily computed
\[
S(\bar{\rho})=\log_2 d - (d-2)/d \approx \log_2 d - 1 \,.
\]
Therefore, the difference in entropy of the time-average of the
computational basis state $\ket{\Psi_0}$ encoding a ``yes''-instance
and a ``no''-instance is at least $c/2$ bits, where $c$ is the size of
output register (see Remark~1). This proves the PSPACE-hardness of
estimating the entropy of the time-average as stated in Theorem~2.

Let the output register consist of only one qubit. Obviously, the
reduced state of time-average to the output qubit is $\ket{0}\bra{0}$
for a ``no''-instance. For a ``yes''-instance we know that the output
qubit of the states $\ket{\Psi_j}:=F^j\ket{\Psi_0}$ is $1$ for
$j=d/4,\ldots,3d/4-1$ and $0$ for all other $j$.  To see this, recall
how $V$ works: at step $r_l/2$ the circuit $V$ changes the output
qubit from $0$ to $1$, at step $r_l$ we obtain the initial state with
the output flipped, at step $r_l+r_l/2$ the output is flipped from $1$
to $0$, and finally in step $2r_l$ we obtain the original input state.

Therefore, the occupation probability of the level $\ket{1}$ of
$\bar{\rho}_{\rm out}$ ($\bar{\rho}$ restricted to the output qubit)
is exactly
\begin{eqnarray*}
\bra{1}\bar{\rho}_{\rm out}\ket{1} & = &
\sum_{j=d/4}^{3d/4-1} \bra{\Psi_j}\bar{\rho}\ket{\Psi_j}\,.
\end{eqnarray*}
Let us denote by $P(j)$ the probability to observe the state
$\ket{\Psi_j}$ if we measure the time-average in the computational
basis. The probability $P(j)$ is given by
\begin{eqnarray*}
P(j) & = & 
\bra{\Psi_j}\bar{\rho}\ket{\Psi_j}\\
& = &
\frac{1}{d^2} \sum_{k,l=0}^{d-1} \omega^{(k-l)j}\,
[\lambda_k=\lambda_l] \\
& = &
\frac{1}{d} + \frac{1}{d^2} \sum_{k\neq l}^{d-1} \omega^{(k-l)j}\,
[\lambda_k=\lambda_l]
\end{eqnarray*}
This seen by observing that $\ket{\Psi_j}=\frac{1}{\sqrt{d}}
\sum_{k=0}^{d-1} \omega^{jk}\ket{k}$. Now we make use of the
Diaconis-Shahshahani bound (see Appendix) applied to the cyclic group
$\Z_d$. The characters of $\Z_d$ are given by $\chi_a(j) =
\omega^{aj}$. The coefficients of the Fourier transform of the
probability density $P=(P(0),\ldots,P(d-1))$ are
\begin{eqnarray*}
\hat{P}(a) & = & \sum_{j=0}^d P(j)\chi_a(j) \\
& = &
\frac{1}{d^2} \sum_{k\neq l\, : \lambda_k = \lambda_l}
\sum_{j=0}^{d-1} \omega^{(k-l+a)j}
\end{eqnarray*}
Note that the possible values for $k-l$ are only the even numbers
$2,4,\ldots,d/2$ since we sum over all pairs $(k,l)$ satisfying $k\neq
l$ and the condition $\cos(2\pi k/d)=\cos(2\pi l/d)$ being equivalent
to $k=-j$.

If $a$ is odd, then $k-l+a\not\equiv 0\mod d$. But if $a$ is even,
then there is exactly one pair of $(k,l)$ satisfying above properties
and $k-l+a\equiv 0\mod d$. With these observations we obtain
\begin{equation}
\hat{P}(a)=\left\{
\begin{array}{cc}
0 & \mbox{ if } a \mbox{ odd} \\
\frac{1}{d} & \mbox{ if } a \mbox{ even} \\
\end{array}
\right.
\end{equation}
Now we obtain with Diaconis-Shahshani bound a lower bound on the total
variation distance between $P$ and the uniform distribution $U$ on the
orbit states $F^j\ket{\Psi_0}$
\[
\|P-U\|_{{\rm TV}}^2\le \frac{1}{4}\,(d/2-1)\,\frac{1}{d^2} \le \frac{1}{8d}
\]
Taking the square root and multiplying both by $2$ we obtain
\[
\sum_{j=0}^{d-1} |P(j)-\frac{1}{d}| \le \frac{1}{\sqrt{2d}}\,.
\]
Using the triangle inequality the above bound implies that
\[
\bra{1}\bar{\rho}_{\rm out}\ket{1} =
\bra{1}\bar{\rho}\ket{1}=\sum_{j=d/4}^{3d/4-1} P(j) \in
[\,\frac{1}{2}-\frac{1}{\sqrt{2d}}\,,\frac{1}{2}+\frac{1}{\sqrt{2d}}\,]\,.
\]
(Note that this is also true if we sum over any set containing half of
all possible $j$.)

In summary, we have proved that for a ``yes''-instance the occupation
probability $\bra{1}\bar{\rho}_{\rm out}\ket{1}$ is almost $1/2$,
whereas it is $0$ for a ``no''-instance.

\section{Appendix}
\proof{Lemma~1: Let $\{\ket{k}\}$ be a basis consisting of
eigenvectors of $H$ with corresponding eigenvalues
$\lambda_k$. Express the state $\rho$ in this basis
\[
\rho = \sum_{k,l} \gamma_{kl} \ket{k}\bra{l}\,.
\]
For the time-average we obtain
\begin{eqnarray*}
\bar{\rho} & = & 
\lim_{T\rightarrow\infty} \frac{1}{T}\int_{t=0}^T 
\sum_{k,l=0}^{d-1}
e^{-i(\lambda_k - \lambda_l) t}\,\gamma_{kl} \ket{k}\bra{\l} \\
& = &
\sum_{k,l=0}^{d-1} 
\left( \lim_{T\rightarrow\infty} \frac{1}{T}\int_{t=0}^T 
e^{-i(\lambda_k - \lambda_l) t}\, \right) \gamma_{kl} \ket{k}\bra{\l} \\
& = &
\frac{1}{d} \sum_{k,l=0}^{d-1} [\lambda_k=\lambda_l]\,\gamma_{kl} 
\ket{k}\bra{l} \\
& = &
\sum_{\lambda} P_{\lambda} \rho P_{\lambda}\,,
\end{eqnarray*}
where $[\lambda_k=\lambda_l]$ is $1$ if both eigenvalues are equal and
$0$ otherwise, $\lambda$ are the different eigenvalues of $H$, and
$P_{\lambda}$ are the orthogonal projection onto the corresponding
eigenspaces.}

The proof of the lemma can be found in \cite{Diaconis:88}.
\begin{lemma}[Diaconis-Shahshahani bound]${}$\\
Let $A$ be an arbitrary abelian group, $P$ an arbitrary probability
distribution on $A$ and $U$ the uniform distribution on $A$. The
Diaconis-Shahshahani bound on the total variation distance
\[
\|P-U\|_{{\rm TV}}:=\frac{1}{2}\sum_{g\in A} |P(g)-U(g)|
\] 
is
\begin{equation}
\|P-U\|_{{\rm TV}}^2 \le \frac{1}{4}\sum_{\chi\neq\chi_0}
|\hat{P}(\chi)|^2\,,
\end{equation}
where the sum runs over all non-trivial irreducible characters of $A$
and $\hat{P}$ denotes the Fourier transform of $P$, that is,
$\hat{P}(\chi)=\sum_{g\in A} \chi(a) P(g)$.
\end{lemma}

\subsection*{Acknowledgements} 
We would like to thank Dominik Janzing for proposing us to consider
the problem of estimating the entropy of time-averages and many
interesting discussions.

This work has been supported by grants of the {\it Landesstiftung
Baden-W\"{u}rttemberg} (project ``Kontinuierliche Modelle der
Quanteninformationsverarbeitung'').

\end{document}